# A Comparison of Methods for Player Clustering via Behavioral Telemetry

Anders Drachen
PLAIT Lab/Game Analytics
360 Huntington Avenue
Boston, MA 02115, USA
andersdrachen@gmail.com

Christian Thurau
Game Analytics
Refshalevej 147, 1
1432 Copenhagen, Denmark
christian@gameanalytics.com

Rafet Sifa
University of Bonn/Game Analytics
Refshalevej 147, 1
1432 Copenhagen, Denmark
rafet@gameanalytics.com

Christian Bauckhage
Fraunhofer IAIS and University of Bonn
Schloss BIrlinghoven
53754 Sankt Augustin, Germany
christian.bauckhage@iais.frainhofer.de

## ABSTRACT
The analysis of user behavior in digital games has been aided by the introduction of user telemetry in game development, which provides unprecedented access to quantitative data on user behavior from the installed game clients of the entire population of players. Player behavior telemetry datasets can be exceptionally complex, with features recorded for a varying population of users over a temporal segment that can reach years in duration. Categorization of behaviors, whether through descriptive methods (e.g. segmentation) or unsupervised/supervised learning techniques, is valuable for finding patterns in the behavioral data, and developing profiles that are actionable to game developers. There are numerous methods for unsupervised clustering of user behavior, e.g. k-means/c-means, Non-negative Matrix Factorization, or Principal Component Analysis. Although all yield behavior categorizations, interpretation of the resulting categories in terms of actual play behavior can be difficult if not impossible. In this paper, a range of unsupervised techniques are applied together with Archetypal Analysis to develop behavioral clusters from playtime data of 70,014 *World of Warcraft* players, covering a five year interval. The techniques are evaluated with respect to their ability to develop actionable behavioral profiles from the dataset.

## 1. INTRODUCTION AND BACKGROUND

With respect to user the variety of user-game interactions and game mechanics, contemporary digital games range from the very simple to the very complex. For the AAA-level major commercial titles the tendency is towards increasing complexity, evident in game forms such as Massively Multiplayer Online Games (MMOGs) where hundreds of thousands of entities, objects and not the least players, form a tightly rule-based but complex amalgam of potential or realized interactions [2, 5, 12].The net effect of the trend towards increasing complexity paralleling technological development in digital games is to grant the user, or player, more ways of interacting with the game, which in turn leads to an increase in the space of potential player behaviors. This trend means that evaluating the effectiveness of a game design is increasingly challenging [5, 19, 26].

Within the past few years, behavior analysis in digital games has moved from the laboratory to the wild thanks to telemetry tracking, and rapidly become a widely applied factor in game development. There are several drivers behind this development. Two of these are, on one hand, the fact that telemetry tracking provides access to the entire population of players, and, on the other hand, the requirement of increasingly popular new business models like Free-to-Play (F2P) for constant monitoring and evaluation of the player population behavior in order to drive revenue [5, 12, 19, 26, 29]. Irrespective of the underlying motivations, user-oriented business intelligence practices have become rapidly assimilated by the game industry.

*User telemetry* is quantitative data about player-game or player-player interaction and is compiled in databases from logs provided from each game client. Raw telemetry data are translated into metrics, e.g. total playtime per user, daily active users, average revenue per user etc. [26]. Any action the player (user) takes while playing can potentially be recorded and stored. The approach forms a strong supplement to the practices of usability- and playability-testing, because user telemetry can provide the kind of detailed quantitative information on player behavior which is excessively time-consuming and in sometimes impossible to obtain using any other approach. The analysis of player behavior via user telemetry is of interest to the investigation of User Experience (UX) in games (Player Experience, PX), because it provides direct evidence of problems affecting the PX, for example indicating where in a game users have problems progressing or understanding the GUI [5, 9, 12].

In the context of digital games, behavioral analysis can be carried out in numerous ways, and given its novelty in the context of digital games as compared to over two decades of application in e.g. web analytics [10], behavioral analysis in games is still in its infancy.

A form of behavioral analysis seeing widespread use in the industry today is segmentation or categorization [26, 29]. The term categorization is here used as a catch-all for any analytical technique which collapses a high number of users into a few profiles, irrespective of the specific method applied (e.g. segmentation, clustering and classification).

Player categorization provides a means for analyzing complex behavioral datasets and distilling the results into behavioral profiles which can be acted upon to test and refine a game design (or specific parameters of a design), as well as inform monetization strategies [1, 15]. The fundamental approach in is to reduce the dimensionality of a dataset, in order to find the most important variables, and express the underlying patterns in the user behavior as a function of these variables, e.g. via defining behavioral profiles which can be used to test/refine parameters of a game design [1, 15]. Notably in games of a persistent nature where the revenue funnel is based on subscriptions or micro-transactions, it is vital to be able to monitor the behaviors of a player- (user-) community.

In addition, behavioral categorization provides the basis for selecting playtesting participants that cover the behavioral range of the users, which supplements traditional demographic approaches towards segmenting target audience for user-oriented testing [19, 9, 12]. Finally, categorization of players based on their behavior forms a key line of investigation driving the research towards the development of adaptive games [25, 26].

In Machine Learning (ML), techniques vary, with examples including unsupervised clustering algorithms (e.g. k-means, c-means, Ward's Linkage) and Non-negative Matrix Factorization (NMF) [14], Self-Organizing Networks (SOM)) [7, 1]. Different approaches have varied strengths and weaknesses, which can make it difficult for non-experts to decide upon a strategy for obtaining patterns from behavioral data [26, 27]. In this paper, a range of commonly used clustering algorithms are compared and evaluated (k-means, c-means, NMF and PCA), and Archetypal Analysis (AA) [4], via the Simplex Volume Maximization (SIVM) algorithm, is introduced as a means for unsupervised behavior categorization in game metrics datasets as a supplement to centroid-seeking techniques.

Archetypal Analysis has only recently been developed to a point where it can handle large-scale data [22]. Unlike cluster-centroid seeking algorithms, AA specifically focuses on players residing on the convex hull of a distribution, i.e. it looks for extreme behaviors. This means that the resulting clusters do not represent users that reside within dense cluster regions [22]. The individual player can via AA be described in terms of a combination of the archetypical profiles generated.

## 2. RESULTS AND SIGNIFICANCE

In this paper Archetype Analysis is compared with a selection of traditionally used methods with the purpose of clustering behavioral data and evaluate the pros and cons of these in developing behavioral profiles that are actionable to game designers. Evaluation is done on **playtime** and **leveling speed** telemetry data recorded from the highly popular Massively Multi-Player Online Role Playing Game (MMORPG) *World of Warcraft* (Figure 1).

Playtime is one of the most widely utilized measures of player behavior in games because it provides a top-down proxy measure of the overall engagement that a player experiences with a game [26] - i.e. the shorter the total playtime, the less the appeal of the game. Optimizing playtime is one of the key challenges of game design, notably for games relying on F2P and similar revenue models [26, 29]. Playtime and its derivatives forms the basis for a number of important Key Performance Indicators (KPIs) in the industry, including Daily Active Users (DAU) and Completion Time (CT), and is essential for calculating user attrition [29].

In terms of results, PCA and NMF generated profiles that are not interpretable in terms of the game´s mechanics, e.g. players decreased in level over time, which is not possible in *World of Warcraft*. In contrast, k-means and AA produced interpretable behavioral profiles (interpretable factorization). However, the k-means centroids are overall similar, showing a similar leveling speed profile across centroids, and do easily allow a straight forward labeling. The AA basis vectors are comparatively more varied, interpretable and can similar to the k-means centroids be used to make assumptions about the leveling behavior of the players. The results thus highlight the pros and cons of centroid-seeking vs. convex-hull seeking unsupervised clustering methods in exploring behavioral telemetry datasets from digital games.

## 3. RELATED WORK

Previous and current work on analytics in the context of digital games is generally divided into two branches: industrial R&D, which is focused on design evaluation, user research and monetization, and academic research, which is focused on the same topics as well as game AI, adaptive games and visualization. To an increasing degree, these overlap as collaborations between game companies and research institutions have become initiated in Europe, Asia and North America. Within this larger context, user profiling via data mining is a relatively new endeavor in game development and –research [26]. Within academic research, the majority of previous work on user behavior analysis has focused on small sample sizes due to the lack of available large-scale datasets but also due to computational constraints. Additionally, research has (mainly) been focused on simple test-bed games, which are inherently simpler than AAA-level commercial games. This is changing with increased collaboration between companies and research institutions, and the available literature is growing, e.g. [26-30] Work on large-scale data in the context of commercial games remains however rare, and highlights the need for robust algorithms to categorize (segment, cluster or classify) data which can scale to large datasets.

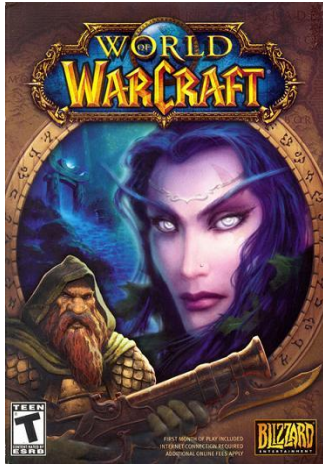

**Fig. 1:** *World of Warcraft*, **one of the original two box covers (Blizzard Entertainment, 2003).**

Clustering algorithms have been explored as the basis for categorizing player behavior, e.g. by Missura and Gärtner [17], who used k-means clustering and support vector machines to predict dynamic difficulty adjustments for a simple shooter-type digital game. Thawonmas and Iizuka [21] adopted frequency analysis to establish patterns of behavior among the player base of the MMOG *Cabal Online*, attempting to identify aberrant behavior which could indicate whether a specific character in the game was controlled by a program (a "bot"), usually developed to mine in-game resources (e.g. gold farming), or a human player. Drachen et al. [30] applied k-means clustering and simplex volume maximization to two datasets covering 260,000 players from the MMORPG *Tera Online* and the multi-player online First-Person Shooter (FPS) *Battlefield 2: Bad Company 2*. Adopting the approach of Drachen et al. [1], they developed behavioral profiles using game design language based on the clusters generated by the two algorithms. Related work on behavioral analysis in games includes Ducheneaut & Moore [6], who utilized action frequencies in the MMOG *Star Wars Galaxies* to categorize player behaviors, focusing on a small set of player-player interaction behaviors, two locations in the game world from a single server (of 25 at the time), 5,493 players and a limited time period (26 days) which is short compared to later longitudinal studies of MMOGs such as Shim and Srivastava [3, 34]. Larger-scale research based on data from commercial games include Weber and Mateas [24], who utilized a series of classification algorithms for recognizing player strategy in over 5000 replays of the Real-Time Strategy (RTS) *StarCraft*, employing regression algorithms in order to predict when specific unit or building types would be produced. Similarly, Thurau and Bauckhage [3], showed how large-scale variants of Archetypal Analysis could be used to analyze the evolution of *World of Warcraft* guilds. Drachen and Canossa [5] employed Geographical Information Systems to provide analyses of spatial behavior for the game *Tomb Raider: Underworld*, showing how such analyses can assist game design. Furthermore, based on a selection of metrics from key game mechanics, Drachen et al. [1] classified the behavior of 1365 players of *Tomb Raider: Underworld*, using Self-Organizing Networks and k-means clustering, locating four classes of user behavior that encompassed over 90 percent of the users in the dataset. These were translated into design terminology for use by the game's developers, Crystal Dynamics. In a follow-up paper, Mahlman et al. [15] classified a sample of 10,000 players using eight categories of game metrics across more than 70 sub-sections of the game. The authors also demonstrated how behavior can be predicted based on analysis of early play profiles, a topic which has since gained substantial traction in the industry where prediction of user behavior is of interest to prevent churn (attrition) and improve monetization [29]. Similarly focused on prediction, Thawonmas et al. [20] studied revisitations in online games focusing on either revisitation to a game or a game area, working with online access logs from 50,000 characters from the MMOG *Shen Zhou Online*, and 60,000 characters from *World of Warcraft*. The results indicated that it is possible to predict revisitations and patterns in these based on behavioral data from digital games.

In summary, the majority of current academic work on player behavior analysis is with few exceptions based on simple test-bed games, behavioral telemetry mined via monitoring the information stream between client and game server in MMORGs [e.g. 35], or scraping online repositories of game data [e.g. 30] – as is the case here - It is more rare to see research based on behavioral telemetry logged in-house by the companies, mainly due to confidentiality issues [1], although exceptions exist, notably [1, 27, 30, 36].

The current state-of-the-art of user categorization analysis – and other analytics work – in the game industry is difficult to evaluate because telemetry data and analysis techniques are considered proprietary. Through industry events such as the Game Developers Conference [e.g. 33], industry magazines, blogs and analytics middleware providers, some knowledge is shared but rarely at the level of detail of academic research - e.g. algorithms used are not revealed. The two books that have recently been published on game analytics [26, 29] provide more detailed insights, indicating that segmentation is generally performed based on pre-defined classes, e.g. "whales", "dolphins" and "minnows" depending on how much money the user in question spends on virtual items or upgrades [29]. This kind of approach using pre-defined features and classes can be useful, but has the inherent problem of fitting data to classes that may not exist in the dataset. Relying on pre-defined classes (or categories) is also susceptible to changes in the player population over time, and prevents exploratory behavior analysis and thus is at risk of missing important patterns in behavioral data.

In summary, the methods adopted for investigating player behavior are varied, but there is a lack of comparative analyses, which forms the main motivation for the analysis presented here. Furthermore, there is a general lack of research in data interpretability, which is notably important in a practical development context, where the results of a classification analysis should be as easy as possible to interpret. Considering the often massive size and increasing complexity of game telemetry datasets, this is an important knowledge gap [23, 26].

## 4. THE IMPORTANCEOF INTERPRETABLE CLUSTERS

The goal of user behavior analysis in game business intelligence is an interpretable representation of the data at hand and the patterns residing in the data, as the representation basically has to assist a human in analyzing huge amounts of game data [e.g. 26-28]. Focusing on clustering, i.e. the unsupervised development of behavioral profiles, ideally, one could assign a simple expressive label to each found basis vector or centroid (each player profile), which can be interpreted by the audience of analysis reports, e.g. game designers [Drachen et al., 2012; Seifl El-Nasr et al., 2013].

While there is no objective criterion on what a descriptive representation is, a key criterion is that results, in order to be interpretable, should embed the data in a lower dimensional space than originally (which is the goal of clustering), and ideally the basis vectors $W$ should correspond to actual data points (actual players). This is the case for the SIVM algorithm as the archetypes the method produces are restricted to being sparse mixtures of individual data points. This makes the method interesting as a means for player clustering because it does not require expert knowledge to interpret the results. This contrasts with other dimensionality reduction methods such as PCA [11] where the resulting elements can lack physical meaning; and NMF which yields characteristic parts, rather than archetypal composites [7]. K-means clustering see the basis vectors residing within cluster regions of the data samples, however, the centroids do not necessarily have to reside on existing data samples. Looking at this from a more philosophical standpoint, searching for extremal points accommodates human cognition, since memorable insights and experiences typically occur in form of extremes rather than as averages. Philosophers and Psychologists have noted this for long, since explanations of the world in terms of archetypes date back to Plato. According to C.G. Jung, it is the opposition that creates imagination. Every wish immediately suggests its opposite and in order to have a concept of good, there must be a concept of bad, just as there cannot be an idea of up without a concept of down. This principle of opposites is best summarized by Hegel's statement that "*everything carries with it its own negation*". The only way we can know anything is by contrast with an opposite. By focusing on extreme opposites, we simply enlarge the margin of what we know and in turn our chance to separate things. In contrast, k-means clustering focuses on the average and can therefore in the context of other centroids be more difficult to interpret. Essentially, while the centroid vectors all cover different regions of the data space, their overall similarity can be so high as to make assigning concrete labels difficult. This contrast between centroid-seeking and convex hull-seeking algorithms for clustering was exemplified in Drachen et al. [30], who notes that techniques like k-means can be useful for gaining insights into the general distribution of behavior in a games´ population, whereas AA is highly suited for detecting extreme behaviors; and by using soft cluster assignments AA provides the ability to build profiles of individual players as a function of their location in the variance space in relation to specific archetypes. For example, a player could be 90% the ultra-fast leveling archetype found here (Figure 5b) and 10% distributed across the other archetypes.

## 5. DATA AND METHOD

In the following, we introduce and discuss the game metrics dataset obtained from *World of Warcraft*, and the clustering methods selected for evaluation and comparison. Specifically, we compare k-means, c-means, NMF, PCA and SIVM (AA). It was found that c-means clustering results were function- ally similar to k-means. Therefore we only include the results of k-means clustering. Our intention here is to investigate how common clustering techniques perform on behavioral telemetry data with respect to (a) descriptive representations, (b) cluster separation. Due to space constraints, a detailed description and evaluation of each of the four sets of results will not be included here, but focus rather applied to the discussion and comparison of the results. The *World of Warcraft* dataset contains, for a set of players, recordings of their online time and their level for a specific date. We aggregate the recordings into a 2.555 dimensional feature vector where each entry corresponds to the level the player reached for each day in the last 6 years. A typical feature vector can be seen in Figure 2. Note that the maximal level of a character was increased twice via expansion packs (from levels 60 to 70 and 70 to 80) during the period of recording (and in December 7th 2010 a third time, following the end of the data logging period, from levels 80 to 85), usually when a new expansion WAS released (evident as stepwise in- creases in several of the profiles in Figure 2).

We applied AA (SIVM), NMF, k-means, c-means and PCA to the dataset. Note that unsupervised methods usually suffer from the problem of having no objective way of defining threshold values, which makes the definition of the number of classes to use a subjective decision. These aspects of cluster analysis add to the difficulty in adopting these methods by non-experts in a game design/development context. A key assumption for all of the evaluated methods is that the game metrics data can be stored in a matrix, s.t. each column corresponds to a particular player or entity. The dataset being used consists of logs of the online playtime and leveling speeds from 70,014 player characters of *World of Warcraft*, randomly selected from a larger dataset comprising approximately 18 million player characters, recorded during eth period from 2005-2010, i.e. several years of playtime. Data for all characters include the time of first appearance in the game. Please note that the dataset is not a complete recording of the players and that we had to interpolate missing values (on the scale of a few percent, mainly due to server outages). Data were obtained from mining the *Warcraft Realms* site. Only two behavioral variables were included here in the interest of clarity in comparing the ability of different algorithms to provide results based on behavioral telemetry.

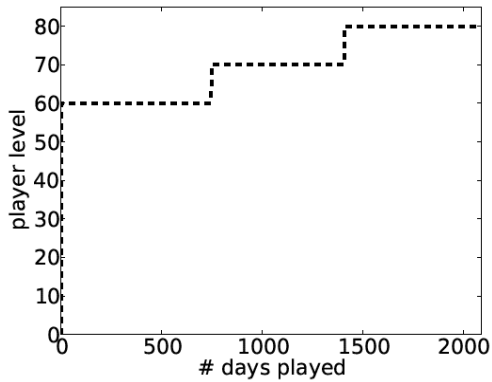

**Fig. 2. Level/time plot of a specific player (in Figures 3-6 basis vectors). The x-axis shows a timeline, the y-axis indicates the current level of the player.**

### 5.1 Behavior analysis by clustering

Clustering is arguably one of the most common steps in unsupervised behavior analysis. Expressing the *d*-dimensional data-set with *n* samples as a column matrix $V^{d \times n}$, the main goal of clustering is to find a set of *k* basis vectors expressed as $W^{d \times k}$ for $k \ll n$. The belongingness values of the data points to each centroids is defined as a coefficient matrix $H^{k \times n}$. Clustering can be interpreted as matrix factorization in terms of finding matrices with lower ranks to approximate the data-sets. This can be shown as finding matrices $W$ and $H$ such that the Frobenius norm $\|V - WH\|$ is minimized. Common approaches to achieve such a factorization include Principal Component Analysis (PCA) [11, 8], k-means clustering (k-means), or Non-negative Matrix Factorization (NMF) [18, 14] and more recently Archetypal Analysis [3, 4, 31]. It is important to note that while all these methods try to minimize the same criterion, they impose different constraints and thus yield different matrix factors. For example, PCA constrains $W$ to be composed of orthonormal vectors and produces a dense $H$, k-means clustering constrains $H$ to unary vectors, and NMF assumes $V$, $W$, and $H$ to be non-negative matrices and often leads to sparse representations of the data.

### 5.2 Archetypal Analysis

Archetypal Analysis (AA) is a soft clustering technique introduced by Cutler and Breiman [33] and mainly aims to describe the dataset as mixture of extreme values called the archetypes. The archetypes are placed in the data convex-hull and increasing the number of archetypes results in a better approximation of the data convex hull [31, 32]. Figure 3 illustrates an example of running k-means and Archetypal Analysis on a two dimensional toy dataset where the black diamonds represent the basis vectors and the grouping for AA is made based on the maximum belongingness value. It is important to note that while the basis-vectors for k-means are centrally located, the archetypes found by Archetypal Analysis are extreme points.

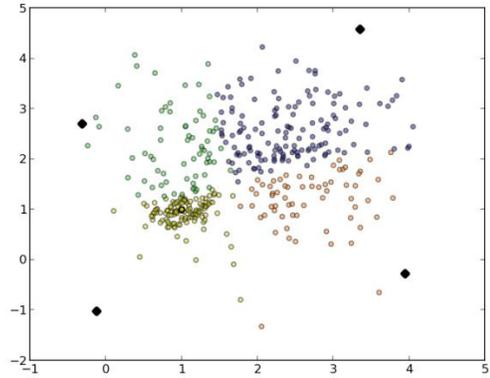

(a) Archetypal Analysis with Simplex Volume Maximization

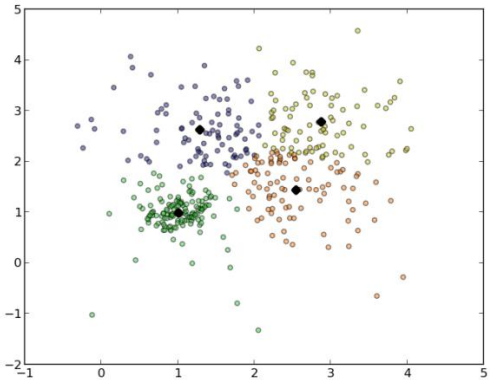

(b) k-means

**Fig. 3: Hard Clustering on a Toy Dataset with 4 Basis Vectors**

We can define the Frobenius norm to be minimized as in Equation (1):

$$E = \|V - VGH\|^2 = \|V - WH\|^2. \quad (1)$$

Where the matrices are defined as $G \in \mathbb{R}^{k \times n}$ and $H \in \mathbb{R}^{k \times n}$ and $W = V G$. Cutler and Breiman [33] presented an alternating constrained least squares algorithm to minimize the Frobenius norm where $G$ is restricted to convexity to describe the archetypes as being convex mixtures of the data entities. This method does not scale well with large data sets [31]. In order to deal with large data sets Thurau et. al. [31] proposed the Simplex Volume Maximization (SIVM) algorithm that calculates the archetypes by fitting a simplex with the maximum volume to the data. SIVM constrains *G* to be composed of unary column vectors which means that the archetypes are selected from the data matrix *V*. The algorithm is shown in Algorithm (1). The main idea of the algorithm is derived from the fact that adding a new basis vector from the dataset to a simplex created from data samples never increases the Frobenius norm of the convex approximation of the data. Instead of minimizing the Frobenius norm, the algorithm iteratively finds vertices maximizing the Cayley-Menger Determinant, which is equal to the volume of a simplex or polytope [31]. Namely, given a simplex *S* with k-1 vertices, the algorithm iteratively finds a new vertex such that: $\omega_i = argmax\ Volume(S \cup x_q)$. Additionally, this is equivalent to finding the vertex $\omega_i$ such that (Equation (2)):

$$\omega_i = argmax\ [\sum_{j=r+1}^{n} d_{rq}[a + \sum_{j=r+1}^{n} d_{jq} - \frac{n-1}{2}\sum_{b=1}^{n} d_{bq}^2]]$$

Where *a* is the length of the edges of the simplex and $d_{ij}$ is the distance between *i*th and *j*th data points in *V*. A Python implementation of the two methods is available from pymf.googlecode.com.

**Algorithm 1** Simplex Volume Maximization

Select $x_i$ randomly from $X$
Choose the first basis vector: $w_1 = \underset{l}{argmax}\ dist(x_l, \underset{z}{argmax}\ dist(x_i, x_z))$
**for** index $i \in [2, k]$ **do**
    Let $S$ be simplex with $i - 1$ vertices.
    Find the maximizer vertex: $w_i = \underset{q}{argmax}\ Volume(S \cup x_q)$
**end for**

## 6. RESULTS

For the presented experiments the number of basis vectors/classes was set to *k* = 8 across all algorithms irrespective of the specific solution offered, based on a consideration of variance explained vs. retaining a useful number of basis vectors with respect to the end goal being to produce player classes that are significantly different behaviorally (Figure 4). The resulting basis vectors or cluster centroids are visualized in Figure 5, 6, 7 and 8. For example, Figure 5(a) shows the level/time history plot of a specific player who only very slowly increased his experience level from level 10 to level 20, and Figure 5(b) shows a player who quickly increased his level to 70, and then after some time to level 80. These two player types can be immediately labeled as "casual player" and "hardcore player". Comparing the resulting basis vectors of the different methods shows that only for k-means clustering (and c-means) and AA (SIVM) did we obtain an interpretable factorization, i.e. a factorization that corresponds with legal behaviors afforded by the design of *World of Warcraft*. The basis vectors of PCA and NMF (Figure 6,7) are not or only partly interpretable. The k-means centroids (Figure 6) display some similarity, i.e. a similar convex shape (with one exception, Figure 6c), with slightly varying slopes. This means that the resulting behavioral profiles, expressed in terms of playtime and leveling speed, will be somewhat similar. In contrast, the SIVM basis vectors in Figure 5 exhibit greater differences and all correspond to legal behaviors (e.g. no gradual increase in character level – this is an instantaneous process). This showcases the benefit of using techniques that produce centroids or basis vectors that correspond to actual player profiles. From the SIVM basis vectors, we can make assumptions about the leveling behavior of the players as the steepest increase in the level seems to correlate with the release of expansion packs (as outlined above) and the simultaneous increase of the maximal level. Besides a descriptive representation of the clusters developed, a quantitative discrimination of player types is desirable - i.e. how many players that belong to each behavioral class. This, however, is only fully supported for k-means clustering as it is the only method that builds on hard cluster assignments, with each sample belonging to only one particular cluster. The other methods are usually soft (or more precisely linear, convex, or non-negative) combinations of their basis vectors. This means that players are expressed in terms of their relationship to each of the eight behavioral profiles (basis vectors) located, and summarily grouped (clustered) according to their distribution in the space spanned by the basis vectors. For the numbers of players belonging to each basis vector provided here, players have been assigned to the nearest basis vector (behavioral profile). This provides profile divisions; however, a more precise way of grouping players would be to define clusters in the space extended by the eight basis vectors. The results indicate that the distribution of players to eight basis vectors across the four methods included (Figure 4) are not similar, with AA and PCA indicating three large groups, and k-means/NMF a division into four large and four smaller groups each.

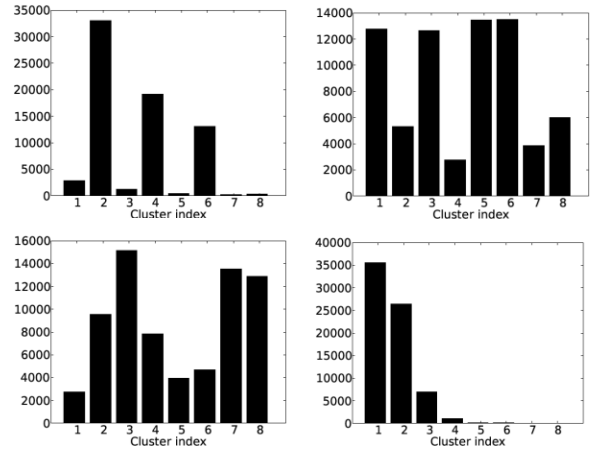

**Fig. 4. Hard assignment of data samples to cluster centroids for (from top left and clockwise) AA, k-means, PCA and NMF, highlighting that the solutions generated by the four algorithms varies substantially.**

## 7. CONCLUSIONS AND DISCUSSION

In the above four different commonly used methods for clustering data derived from humans and human behavior (k-means, c-means, NMF, PCA) have been applied with the purpose of defining classes of player behavior based on a game metrics dataset from the MMOG *World of Warcraft* using two behavioral variables: playtime and leveling speed.

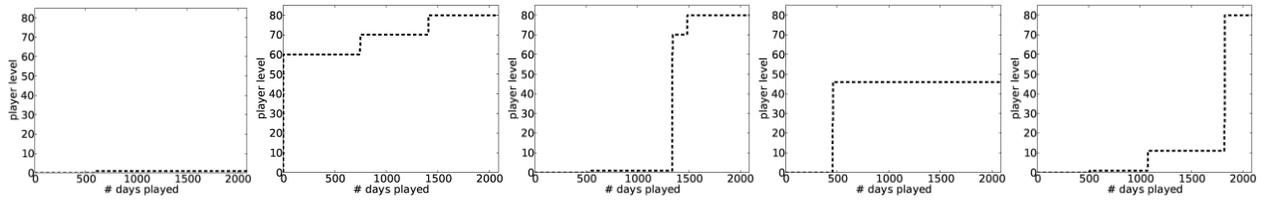

**Fig. 5** The first five basis vectors for Archetypal Analysis. These reside on actual data samples (players in this case). All basis vectors correspond to legal player behavior (e.g. players do not loose levels). Note the straight line segments which map directly to level increases. Archetypes are often pairwise polar opposites, which further supports interpretability.

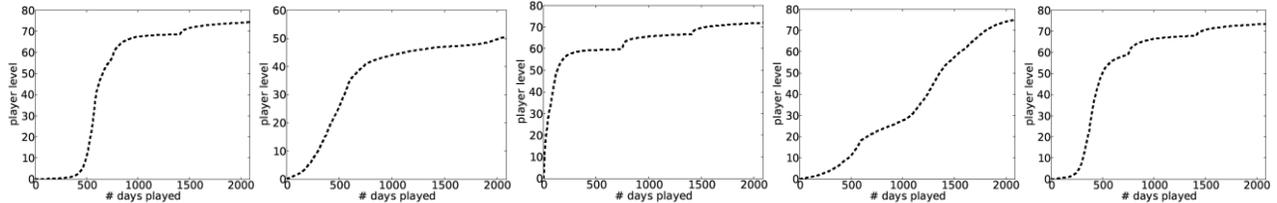

**Fig. 6** The first five cluster centroids for k-means clustering reside on center locations of cluster regions. While they accurately represent a broad number of players, they are overall very similar to each other and do not allow straight forward interpretation.

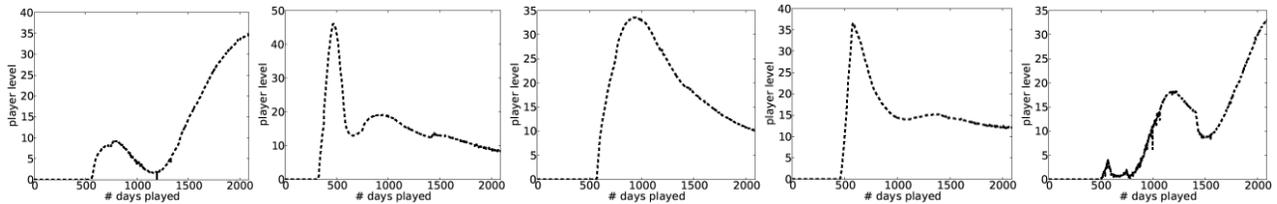

**Fig. 7** The first five basis vectors for non-negative matrix factorization represent parts of original data samples. As they are stricly positive, they allow for interpretation but they do not correspond to actually existing players or behaviors that are possible in the game, e.g. characters lose levels. .

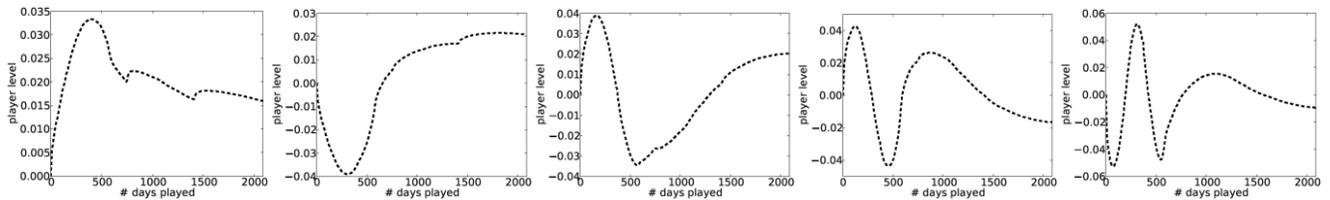

**Fig. 8** The first five basis vectors for principal component analysis do not correspond to actual players, and also correspond to behaviors that are not possible in the game, e.g. loss of character levels.

The effectiveness of these methods previously employed to analyze behavioral telemetry datasets from digital games, have been compared with Archetypal Analysis (AA), via Simplex Volume Maximization, which has only recently been adopted for use with large-scale datasets [3, 30, 31]. The results indicate that different approaches towards classifying player behavior in digital games have different strengths and weaknesses. K-means/c-means clustering allocate players directly to groups, via cluster centroids, that are defined by specific behaviors, whereas NMF, PCA and AA initially provide basis vectors which span a space that players fall within, which means that players can be described in terms of their relationship with each basis vector ("behavior type") and grouped (clustered) according to their distribution in the space spanned by the basis vectors. While this might appear to make clustering more attractive than the other methods as it saves a step in the analysis, there is an important drawback: the clusters tend towards being similar and thus not conducive to interpretation in terms of differences in the behavior of the different clusters of players. A similar interpretative problem is evident for NMF and PCA, which provide results that are counterintuitive to the underlying behaviors. For example, players can be seen to loose levels, something that is not possible in *World of Warcraft* (Figure 7, Figure 8). While NMF and PCA may provide valid basis vectors from a methodological perspective, based on the available behavioral data, these are not intuitively interpretable in terms of the behavior of the players. In comparison, AA and k-means/c-means (Figure 5, Figure 6) provide basis vectors that are intuitively interpretable from the perspective of the underlying behaviors they signify. Only AA; however, has in the presented case resulted in basis vectors (archetypical behaviors) that are significantly different. While based on playtime and leveling speed data from a single game, the results presented indicate that AA is a potentially useful method for the classification of player behavior in games, presenting intuitively interpretable behavioral profiles. The results presented indicate that method choice impacts result in large-scale cluster analysis, and highlights the challenges faced by the game industry looking to evaluate user behavior at the scale that has become possible with the introduction of game metrics logging in game development. There is a clear need for research addressing issues such as scaling effects, data types and methodologies for analysis. This is in addition to the overall question of how to relate behavior to experience measures.

# REFERENCES

[1] A. Drachen, G. N. Yannakakis, A. Canossa and J. Togelius. Player Modeling using Self-Organization in Tomb Raider: Underworld. *In Proc. of IEEE CIG,* 2009.
[2] J. Bohannon. Game-Miners Grapple With Massive Data. *Science*, 330(6000):30-31, 2010.
[3] C.Thurau and C. Bauckhage. Analyzing the evolution of social groups in world of warcraft. In *Proc. of IEEE CIG*, 2010.
[4] A. Cutler and L. Breiman. Archetypal Analysis. *Technometrics*, 36(4):338-347, 1994.
[5] A. Drachen and A. Canossa. Evaluating motion. Spatial user behavior in virtual environments. *Int. Journal of Arts and Technology*, v. 4 N3, 2011.
[6] N. Ducheneaut and R. J. Moore. The Social Side of Gaming: A study of interaction patterns in a Massively Multiplayer Online Game. In *Proc. of the 2004 ACM Conf. on Computer supported cooperative work*, 2004.
[7] L. Finesso and P. Spreij. Approximate Nonnegative Matrix Factorization via Alternating Minimization. In *Proc. 16th Int. SMTNS*, 2004.
[8] G. Golub and J. van Loan. *Matrix Computations*. Johns Hopkins University Press, 3rd edition, 1996.
[9] K. Isbister and N. Schafer. *Game Usability*. MK Pub., 2008.
[10] B. J. Jansen. In *Understanding User-Web Interactions via Web Analytics*. Morgan & Claypool Publishers, 2009.
[11] I. Jollie. *Principal Component Analysis*. Springer, 1986.
[12] J. H. Kim, D. V. Gunn, E. Schuh, B. C. Phillips, R. J. Pagulayan, and D. Wixon. Tracking real-time user experience (true): A comprehensive instrumentation solution for complex systems. In *Proceedings of CHI*, 2008.
[13] D. King and S. Chen. Metrics for Social Games. In *Proceedings of the Social Games Summit,* 2009.
[14] D. D. Lee and H. S. Seung. Learning the Parts of Objects by Non-negative Matrix Factorization. *Nature*, 401(6755):788-799, 1999.
[15] T. Mahlman, A. Drachen, A. Canossa, J. Togelius, and G. N. Yannakakis. Predicting Player Behavior in Tomb Raider: Underworld. In *Proceedings of IEEE Computational Intelligence in Game*s, 2010.
[16] L. Mellon. *Applying metrics driven development to MMO costs and risks.* Versant Corporation, 2009.
[17] O. Missura and T. Gärtner. Player modeling for intelligent difficulty adjustment. In *Proc. of the ECML-09 Workshop From Local Patterns to Global Models*, 2009.
[18] P. Paatero and U. Tapper. Positive Matrix Factorization: A Non-negative Factor Model with Optimal Utilization of Error Estimates of Data Values. *Environmetrics*, 5(2):111-126, 1994.
[19] R. Pagulayan, K. Keeker, D. Wixon, R. L. Romero, and T. Fuller. User-centered design in games. In *The Human-Computer Interaction Handbook: Fundamentals, Evolving Technologies, and Emerging Applications*, 883-903. L. Erlbaum Associates, 2003.
[20] J.-K. L. R. Thawonmas, K. Yoshida and K.-T. Chen. Analysis of revisitations in online games. *Jour. of Ent. Comp.*, 2011.
[21] R. Thawonmas and K. Iizuka. Visualization of online game players based on their action behaviors. *Int. Journal of Computer Games Technology*, 2008.
[22] C. Thurau, K. Kersting, and C. Bauckhage. Convex Non-Negative Matrix Factorization in the Wild. In *Proc. IEEE Int. Conf. on Data Mining*, 2009.
[23] C. Thurau, K. Kersting, M. Wahabzada, and C. Bauckhage. Descriptive matrix factorization for sustainability: Adopting the principle of opposites. *Journal of Data Mining and Knowledge Discovery*, 2011.
[24] B. Weber and M. Mateas. A Data Mining Approach to Strategy Prediction. In *Proc. IEEE Symposium on CIG*, 2009.
[25] G. N. Yannakakis and J. Hallam. Real-time Game Adaptation for Optimizing Player Satisfaction. *IEEE Trans. on Computational Intl. and AI in Games*, 1(2):121-133, 2009.
[26] Seif El-Nasr, M.; Drachen, A. and Canossa, A.: *Game Analytics – Maximizing the Value of Player Data*. Springer Publishers, 2013.
[27] B. Medler. *Play with data – an exploration of play analytics and its effect on player experiences*. PhD-Thesis, School of Literature, Communication and Culture, Georgia Tech.
[28] G. N. Yannakakis. Game AI Revisited. In *Proc. of ACM Computing Frontiers Conference,* 2012.
[29] Fields, T. and Cotton, B. *Social Game Design: Monetization Methods and Mechanics*. MK Pub., 2011.
[30] Drachen, A.; Sifa, R.; Bauckhage, C. and Thurau, C.: Guns, Swords and Data: Clustering of Player Behavior in Computer Games in the Wild. In *Proc. IEEE CIG 2012*, 163-170.
[31] C. Thurau, K. Kersting and C. Bauckhage. Yes We Can: Simplex Volume Maximization for Descriptive Web-Scale Matrix Factorization. In *Proc. 19th ICIKM*, 2010, 1785-1788.
[32] C. Bauckhage and C. Thurau. Making archetypal analysis practical. *Pattern Recognition* (2009): 272-281.
[33] G. Zoeller. *Game Development Telemetry*. Presentation at the Game Developers Conference, 2011.
[34] K. J. Shim and J. Srivastava. Behavioral Profiles of Character Types in EverQuest II. In *Proceedings of the IEEE Conference on Computational Intelligence and Games* (CIG-10), 2010.
[35] Y.-T. Lee, C. Kuan-Ta, Y.-M. Cheng and C.-L. Lei. World of Warcraft Avatar History Dataset. In *Proc. of MMSys´11,* 2011.
[36] M. Seif El-Nasr, B. Aghabeigi, D. Milam, M. Erfani, B. Lameman, H. Maygoli and S. Mah. Understanding and Evaluating Cooperative Games. In *Proc. C HI 2010*, 253-262.